\documentclass[preprintnumbers, floatfix, twocolumn,
preprintnumbers, PRD, superscriptaddress,nofootinbib]{revtex4-1}
\usepackage{CJK}
\usepackage{amsmath,amssymb,bm,mathrsfs}
\usepackage{mathtools,mathptmx,array,slashed}
\usepackage{eurosym}
\usepackage{amsfonts}
\usepackage{amsmath}
\usepackage{amssymb,epsf}
\usepackage{latexsym}
\usepackage{graphicx,epsfig}
\usepackage{amssymb}
\usepackage{subfigure}
\usepackage{epstopdf}
\usepackage[colorlinks=true,citecolor=blue,linkcolor=blue,urlcolor=black]{hyperref}
\usepackage{color}
\usepackage{float}

\graphicspath{{Images/}}

\renewcommand{\&}{\textup{\symbol{`\&}}}
\usepackage{hyperref}
\usepackage{amsfonts}
\usepackage{amsmath}

\usepackage{amssymb}
\usepackage{graphicx}%
\usepackage{bigints}
\usepackage{graphicx}
\usepackage{subfigure}
\usepackage{epsf}
\usepackage{bm}
\usepackage{amsmath}
\usepackage{amsfonts}
\usepackage{amssymb}
\usepackage{graphicx}
\usepackage{tabularx}
\usepackage{color}%
\providecommand{\U}[1]{\protect\rule{.1in}{.1in}}
\newcommand{\be}{\begin{equation}}
\newcommand{\ee}{\end{equation}}

\newcommand{\mincir}{\raise
-3.truept\hbox{\rlap{\hbox{$\sim$}}\raise4.truept\hbox{$<$}\ }}
\newcommand{\magcir}{\raise
-3.truept\hbox{\rlap{\hbox{$\sim$}}\raise4.truept\hbox{$>$}\ }}

\def\be{\begin{equation}}
\def\ee{\end{equation}}
\def\bea{\begin{eqnarray}}
\def\eea{\end{eqnarray}}
\def\ba{\begin{aligned}}
\def\ea{\end{aligned}}

\providecommand{\U}[1]{\protect\rule{.1in}{.1in}}

\usepackage{tikz,xcolor,hyperref}

\begin{document}
\title{ Apparent Dark Matter Inspired by Einstein Equation of State}

 \author{Kimet Jusufi}
\email{kimet.jusufi@unite.edu.mk}
\affiliation{Physics Department, State University of Tetovo, Ilinden Street nn, 1200, Tetovo, North Macedonia}
\author{Ahmad Sheykhi}
\email{asheykhi@shirazu.ac.ir} \affiliation{Physics Department and
Biruni Observatory, College of Science, Shiraz University, Shiraz
71454, Iran}

\begin{abstract}
The purpose of this article is twofold. First, by means of
Padmanabhan's proposal on the emergence nature of gravity, we
recover the $\Lambda$CDM model and the effect of the dark matter
in the context of cosmology. Toward this goal, we use the key idea
of Padmanabhan that states cosmic space emerges as the cosmic time
progress and link the emergence of space to the difference between
the number of degrees of freedom on the boundary and in the bulk.
Interestingly enough, we show that the effect of the cold dark
matter in the cosmological setup can be understood by assuming an
interaction between the numbers of degrees of freedom in the bulk.
In the second part, we follow the Jacobson's argument and obtain
the modified Einstein field equations with additional dark matter
component emerging due to the interaction term between dark energy
and baryonic matter related by $\Omega_{DM,0}=\sqrt{2 \alpha
\Omega_{M,0} \Omega_{DE,0}}$, where $\alpha$ is a coupling
constant. Finally,  a correspondence with Yukawa cosmology is
pointed out, and the role of massive gravitons as a possibility in
explaining the nature of the dark sector as well as the
theoretical origin of the Modified Newtonian Dynamics (MOND) are
addressed. We speculate that the interaction coupling $\alpha$
fundamentally measures the entanglement between the gravitons and
matter fields and there exists a fundamental limitation in
measuring the gravitons wavelength.
\end{abstract}

\maketitle

%%%%%%%%%%%%%%%%%%%%%%%%%%%%%%%%%%%%%%%%%%%%%%%%%%%%%%%%%%%%%%%%%%%%%%%%%%%%%%%%%%
\section{Introduction}
After discovery of the black holes thermodynamics in 1970's,
physicists have been speculating that there should be a
correspondence between the laws of gravity and the laws of
thermodynamics. In 1995, Jacobson \cite{Jac} disclosed that the
Einstein's field equations of gravity can be derived by applying
the first law of thermodynamics, $\delta Q=T\delta S$, on the
boundary of spacetime, where $\delta Q$ is the energy flux across
the boundary, $T$ and $S$ are, temperature and entropy associated
with the boundary, respectively. This derivation is of great
importance and reveals that the first law of thermodynamics, when
is applied to a spacetime as a thermodynamic system at large
scales, can be translated to the field equations of gravity.
Following Jacobson, a lot of works have been done to address the
correspondence between the first law of thermodynamics and the
field equations of gravity in different setups
\cite{Pad1,Pad2,Pad3,CaiKim,Cai2,wang1,SheyWC,SheyWC2,SheyCQ,SheyLog,SheyPL,DiGennaro:2022grw}.

According to the $\Lambda$CDM cosmological model (which stands as
a prominent and widely accepted framework in describing the
evolution of the Universe), our Universe comprises three essential
constituents: cold dark matter, dark energy, and baryonic matter
\cite{Springel:2006}. Cold dark matter, a substance that eludes
electromagnetic interaction and manifests solely through its
gravitational influence, offers a potential solution to many
problems, including the observed flat rotation curves of spiral
galaxies \cite{Navarro:1997, Navarro:2010, Moore:1999,
Gilmore:2007, Salucci:2007, Naray:2008}. In recent years, dark
matter has been widely believed to be made of particles beyond the
standard model, which describes baryonic matter. However, today,
such an elusive dark matter particle remains undetected. On the
other hand, the cosmological constant gives the dark energy, which
emerges as a manifestation of energy permeating every corner of
the Universe, frequently associated with vacuum energy
\cite{Weinberg:1989}.

As an alternative to dark matter, in addressing the flat rotation
curves, the Modified Newtonian Dynamics (MOND) theory was
formulated by Milgrom \cite{Milgrom1, Milgrom2, Milgrom3} (see
also \cite{Ferreira:2009, Milgrom:2003, Tiret:2007, Kroupa:2010,
Cardone:2011, Richtler:2011, Bekenstein:2004}). It was argued that
the theoretical origin of the MOND theory can be understood from
Debye entropic gravity perspective \cite{Li,SheMOND}. The idea
that gravity is not a fundamental interaction and can be regarded
as an entropic force caused by the changes in the information
associated with the positions of material bodies, was proposed by
Verlinde \cite{Ver}. Verlinde's derivation of Newton's law of
gravitation at the very least offers a strong analogy with a well
understood statistical mechanism. Therefore, this derivation opens
a new window to understand gravity from the first principles. The
study on the entropic  force has raised a lot of attention
recently (see \cite{Cai4,sheyECFE,Ling,Modesto,Yi,Sheykhi2} and
references therein). In the framework of mimetic gravity, the
MOND-like acceleration was recovered \cite{vagnozzi}. According to
the entropic force scenario, dark matter represents an apparent
manifestation, a consequence of baryonic matter
\cite{Verlinde:2016toy}. Using the entropic force, modified
Friedmann equations have been obtained recently
\cite{Jusufi:2022mir, Millano:2023ahb, Jusufi:2023ayv}.

Verlinde's proposal on the entropic nature of gravity opened a new
window on the origin of gravity, however, it assumes the
background geometry as a preexist structure. An important question
is how one can consider the spacetime as an emergent structure?
Padmanabhan \cite{Padmanabhan:2012ik} was the first who answered
this question and argued that the spatial expansion of our
universe can be regarded as the consequence of emergence of space
and \textit{the cosmic space is emergent as the cosmic time
progresses}. By calculating the difference between the surface
degrees of freedom and the bulk degrees of freedom in a region of
space and equating the result with the volume change of the space,
he derived the Friedmann equations describing the evolution of the
universe \cite{Padmanabhan:2012ik}. Following Padmanabhan, a lot
of works have been carried out to investigate the emergence nature
of gravity and various aspects/extension of this theory have been
addressed in the literatures
\cite{Cai1,Yang,FQ,Shey1,Shey2,FF,Eune,Shey3}.

In this paper, we would like to employ a novel idea and address
the dark matter problem in the context of Padmanabhan's proposal
of emergence gravity.  This leads to modification of Einstein
equations of general relativity, too. This paper is structured as
follows. In the next section we recover the $\Lambda$CDM model by
assuming an interaction between baryonic matter and dark energy in
the bulk. In section III, Following Jacobson \cite{Jac}, we
recover the modified Einstein's field equation with an additional
energy-momentum term which can be regarded as the apparent dark
matter. In section IV, we establish a correspondence with Yukawa
cosmology and explain the role of the massive gravitons as a
possible candidate to explain the nature of the dark sector and the emergence of MOND. We
finish with final remarks in section V.

%%%%%%%%%%%%%%%%%%%%%%%%%%%%%%%%%%%%%%%%%%%%%%%%%%%%%%%%%%%%%%%%%%%%%%%%%%%%%%%%%%%%%%%%%%%%
\section{Recovering $\Lambda$CDM in Padmanabhan's Emergent Universe}
\label{sectII} To explain why our Universe is expanding,
Padmanabhan used the holographic principle and the equipartition
law of energy \cite{Padmanabhan:2012ik}. According to Padmanabhan
the basic law governing the emergence of space must relate the
emergence of space to the difference between degrees of freedom on
the surface and in the bulk, namely $(N_{\rm sur} - N_{\rm
bulk})$. He proposed relation \cite{Padmanabhan:2012ik}
\begin{equation}
 \frac{d V}{d t} = L_P^2  (N_{\rm sur} - N_{\rm bulk}),
\label{key1}
\end{equation}
where $V$ and $t$ are, respectively, the volume and the cosmic
time in Planck units. More generally, it is expected $(\Delta
V/\Delta t)$ to be some function of $(N_{\rm sur} - N_{\rm bulk})$
which vanishes when the latter does (note also that $L_P^2=\hbar
G/c^3$). In the present paper, we would like to modify the above
relation as
 \begin{equation} \label{key2}
 \frac{d V}{d t} = L_P^2  \left[N_{\rm sur} - \left(N_{\rm bulk}+N^{\rm int}_{\rm bulk}\right)\right],
\end{equation}
where we assume there is an additional term for the degrees of
freedom in the bulk denoted by $N^{\rm int}_{\rm bulk}$, which
originates from the interaction between matter/energy components
in the bulk. The origin of this interaction term will be discussed
later on. Such a universe obeys the holographic principle where
$N_{\rm sur}$, the number of degrees of freedom on the spherical
surface of Hubble radius $H^{-1}$, given by
\cite{Padmanabhan:2012ik}
\begin{equation} \label{Ns}
 N_{\rm sur} = \frac{4\pi}{L_P^2 H^2}.
\end{equation}
Here $N_{\rm bulk}$ is the effective number of degrees of freedom
which obeys the equipartition law of energy contained inside the
universe \cite{Padmanabhan:2012ik}
\begin{equation}
 N_{\rm bulk}=\frac{|E|}{(1/2) k_BT}.
\end{equation}
We shall assume a multi-fluid matter inside the universe with
proper Komar energy density
\begin{eqnarray}
 E&=&\sum_i (\rho_i +3p_i)V,
\end{eqnarray}
and hence we can write the equipartition law of energy as
\begin{equation} \label{Nb}
 N_{\rm bulk} = -\sum_i \epsilon_i  \frac{2(\rho_i +3p_i)V}{k_BT},
\end{equation}
along with
 \begin{equation} \label{key2}
 \frac{d V}{d t} = L_P^2  \left[N_{\rm sur} - \left(\epsilon_i N_{\rm bulk}+\epsilon^{\rm int}N^{\rm int}_{\rm bulk}\right)\right],
\end{equation}
where, in our definition, $\epsilon_i=-1$ stands for baryonic
matter and radiation and $\epsilon_i=1$ for dark energy [in order to ensure that $N_{\rm bulk}>0$;  for matter and radiation  $\rho^{M}_{i} +3p^{M}_{i}>0$ and $\rho^{R}_{i} +3p^{R}_{i}>0$, and $\rho_{DE} +3p_{DE}<0$ for dark energy, respectively]. Furthermore, we take $\epsilon^{\rm int}=-1$, meaning that $N^{\rm int}_{\rm bulk}>0$, which should behave as baryonic
matter for reasons that we will explain below.
%\textcolor{red}{It seems $\epsilon_i=1$ for matter and radiation
%and $\epsilon_i=-1$  for DE. The reason is that $N_{\rm bulk}>0$.
%Since for DE we have $\rho_{DE} +3p_{DE}<0$, while for matter and
%radiation $\rho_{i} +3p_{i}>0$. }

The volume of the universe and the temperature associated with the
horizon are given by
\begin{eqnarray} \label{VT}
V=\frac{4\pi}{3H^3},\,\,\,\,\,T=\frac{H}{2\pi}.
\end{eqnarray}
 The novel idea is to consider an interaction term between
baryonic matter and dark energy in the bulk. As a result, we
propose there is an additional degrees of freedom which comes from
the interaction term and assume, for pure dimensional reasons, it
can be written as
\begin{eqnarray} \label{Nint}
    N^{\rm int}_{\rm bulk} =-\epsilon^{\rm int}\sqrt{ \alpha  \   N_{\rm bulk}^{\rm M} N_{\rm bulk}^{\rm DE} },
\end{eqnarray}
where $\alpha$ is a coupling constant, $N_{\rm bulk}^{\rm M}$ and
$ N_{\rm bulk}^{\rm DE}$ are the degrees of freedom of  baryonic
matter and dark energy in the bulk, respectively. In the next
section, we will give further physical arguments about $\alpha$.
Since $p_M=0$ for baryonic matter, and $p_{DE}=-\rho_{DE}$ for
dark energy fluid, then, for the interaction term, we get
\begin{eqnarray} \label{Nint2}
   N^{\rm int}_{\rm bulk}=\sqrt{2 \alpha \rho_M \rho_{DE}  } \frac{2\,V}{k_B T}.
\end{eqnarray}
Let us now obtain the standard Friedmann equation. Using Eqs.
(\ref{key2}), (\ref{Ns}), (\ref{Nb}), (\ref{VT}),  (\ref{Nint2})
and $p_i=\omega_i \rho_i$, we get
\begin{eqnarray}\notag
\frac{dV}{dt}&=& 4\pi L_P^2 \Big[ \frac{1}{L_P^2 H^2}+\sum_i
\rho_{i}(1 + 3 \omega_i )  \frac{4 \pi}{3 H^4}\\
&+& \sqrt{2
\alpha \rho_M \rho_{DE}}  \frac{4 \pi}{3 H^4} \Big],
\end{eqnarray}
where we have set $k_B=1$. From the last equation, we obtain
\begin{equation}
 \frac{\ddot a}{a}=-\frac{4\pi L_P^2}{3} \left[ \sum_i \rho_{i}(1 + 3 \omega_i )+\sqrt{2 \alpha \rho_M \rho_{DE}}\right].
\label{frw}
\end{equation}
In the second term, we have an evolving matter given by $\rho_{M}=
\rho_{M,0} (1+z)^3$ and dark energy $\rho_{DE}=\rho_{DE,0}$, where
it is natural to identify this term as the dark matter for reasons
we shall argue below.  Let us define the following equation as an
evolving dark matter
\begin{equation} \label{rhodm}
\frac{\rho_{DM}}{(1+z)^{3/2}}\equiv\sqrt{2 \alpha \rho_M
\rho_{DE}},
\end{equation}
with using $\rho_{M}= \rho_{M,0} (1+z)^3$ and
$\rho_{DE}=\rho_{DE,0}$, leads to
\begin{equation}\label{rhodm2}
\rho_{DM}=\rho_{DM,0} \,a^{-3},
\end{equation}
where $1+z=a^{-1}$ and $ \rho_{DM,0}\equiv\sqrt{2\alpha \rho_{M,0}
\rho_{DE,0}}$. Note that assuming dark matter is pressureless,
then Eq. (\ref{rhodm2}) is also solution of  the continuity
equation $\dot{\rho}_{DM}+3H \rho_{DM}=0$, as expected and
justifies ansatz (\ref{rhodm}).

If we multiply both sides of Eq. (\ref{frw}) by $2 a \dot{a}$, and
using the continuity equation
\begin{equation}\label{Cont_gen}
\dot{\rho}_i+3H(1+ \omega_i) \rho_i=0,
\end{equation}
where $H=\dot{a}/a$ is the Hubble parameter and $\rho_i=\rho_{i,0}
a^{-3(1+\omega_i)}$, assuming several matter fluids with a
constant equation of state parameters $\omega_i$. We then arrive
at
%\textcolor{red}{ I repeat your calculations and I think the minus
%$-$ appears in the r.h.s in the following}
\begin{equation}
 d(\dot{a}^2)= \frac{8\pi L_P^2}{3} d\left( \sum_i  \rho_{i,0} a^{-1-3 \omega_i}+\rho_{DM,0}\,a^{-1-3\omega_{DM}}\right),
\label{2frw}
\end{equation}
where $\Omega_{DM}=0$. In the l.h.s, we get a constant of integration $k$ as follows
\begin{equation}
 \dot{a}^2+k=  \frac{8\pi L_P^2}{3}\int d\left( \sum_i  \rho_{i,0} a^{-1-3 \omega_i}+\rho_{DM,0}a^{-1}\right).
\label{3frw}
\end{equation}
By solving the integral in the r.h.s, we get
\begin{eqnarray}\notag
    \left(\frac{\dot{a}}{a}\right)^2&=& \frac{8\pi L_P^2}{3}\Big(\rho_{R,0}\, a^{-4}+\rho_{M,0}a^{-3}+\rho_{DM,0}a^{-3}\\
    &+&\rho_{\rm cur} a^2+\rho_{DE,0}\,a^0\Big),
    \end{eqnarray}
where  we have also included the spatial curvature density
%\textcolor{red}{I expect in the limiting case where $N^{\rm
%int}_{\rm bulk}=0=\rho_{DM,0}$, we recover the standard Friedmann
%eq. But it does not. Something is wrong!!! The minus on the r.h.s
%does not exist in Standard Friedmann eq}
%\textcolor{blue}{I have revised up to here. After the above
%clarifications I will continue my revision. 23 Jan 2024}
\begin{eqnarray}
    \rho_{\rm cur}\equiv\frac{3 k}{8 \pi L_P^2 a^2}.
\end{eqnarray}
If we use the critical density
\begin{eqnarray}
    \rho_{\rm crit}=\frac{3 H_0^2}{8 \pi L_P^2 },
\end{eqnarray}
we get the equation in terms of the density parameters
or in terms of redshift $z$ as follows
\begin{equation}\label{EofZ1}
  E^2(z)=\Omega_{R,0}(1+z)^4+ \Omega_{M,0}^{\rm tot}(1+z)^{3}
  +\Omega_{\rm curv} (1+z)^2+\Omega_{DE,0},
  \end{equation}
  where  $E(z)=H/H_0$, and
\begin{equation}\label{defofODM0}
\Omega_{DM,0}=\sqrt{2\,\alpha\, \Omega_{M,0} \Omega_{DE,0} },
\end{equation}
along with
\begin{eqnarray}
    \Omega_{M,0}^{\rm tot}=\Omega_{M,0}+\sqrt{2\,\alpha\, \Omega_{M,0} \Omega_{DE,0} }=\Omega_{M,0}+\Omega_{DM,0}.
\end{eqnarray}
That suggests that dark matter can evolve due to the expansion of
the Universe, which is related to the fact that the baryonic
matter evolves, and the coupling parameter is a function of $z$,
as we pointed out. As a special case, if we consider a flat
universe with $k=0$ for the late-time Universe, which is
consistent with the result obtained in \cite{Jusufi:2023xoa}. We
shall elaborate more about this connection in this work. However,
it is important to see from these equations that dark matter is
only an apparent effect due to the interaction between the dark
energy and matter. That is the key distinction as compared to the
$\Lambda$CDM model, whose respective Hubble parameter as a
function of the redshift for a flat space ($\Omega_{\rm curv}=0$)
is given by (written in our notation)
\begin{equation}
    H(z)=H_{0}\sqrt{\Omega_{R,0}(1+z)^{4}+\Omega_{M,0}^{\rm tot}(1+z)^{3}+\Omega_{DE,0}},
\end{equation}
the last expression is obtained from Eq. \eqref{EofZ1} and it
perfectly matches the $\Lambda$CDM model.
%%%%%%%%%%%%%%%%%%%%%%%%%%%%%%%%%%%%%%%%%%%%%%%%%%%%%%%%%%%%%%%%%%%%%%%%%%%%%%%%%%%%%%%%%%%%%%%%%%%%%%
\section{Einstein field equations with apparent dark matter}
In this section, we will closely follow the seminal paper by
Jacobson \cite{Jac}. Specifically, Jacobson derived the Einstein's
field equations of gravity through the application of the first
law of thermodynamics and one can view the Einstein's field
equation as an equation of state. In the current study, we aim to
extend this framework, incorporating the influences of baryonic
matter, dark energy, and an additional term stemming from the
interaction between baryonic matter and dark energy. First, we
need to consider the heat flow across the horizon attributed to
the energy carried by the matter field, characterized by an
energy-momentum tensor $T^M_{ab}$ representing baryonic matter
\begin{equation}
\delta Q^M =-\kappa \int_H \xi T^M_{ab} k^a\,k^b
d\xi dS.
\end{equation}
In this equation with $S$ we represent the area of the horizon
$H$. The vector $k^a$ corresponds to the tangent vector of the
horizon generators, further $\kappa$ is known as the surface
gravity and, finally, $\xi$ is an appropriate affine parameter. We
also introduce a dark energy term contribution
\begin{equation}
\delta Q^{DE} =-\kappa \int_H \xi T^{DE}_{ab} k^ak^b
d\xi dS.
\end{equation}
Building upon the concept introduced in the preceding section,
which serves as the fundamental idea in this paper, we incorporate
an interaction term between baryonic matter and dark energy.
Consequently, we anticipate the emergence of a contribution term
\begin{equation}
\delta Q^{int.} =-\kappa \int_H  \xi T^{int}_{ab}\,
k^a\,k^b d\xi dS,
\end{equation}
where we have introduced an interaction term
\begin{eqnarray}
    T^{int}_{ab}=u_a u_b \sqrt{2 \alpha \rho_{M,0} \rho_{DE,0}}\,\,a^{-3},
\end{eqnarray}
where $\alpha$ is an interaction parameter between dark energy and
baryonic matter, $a$ is the scale factor, here 4-velocity of the
prefect fluid with $g^{ab} u_{a} u_{b}=-1$. For reasons that will
be clarified later on we shall identify the interaction term as
the dark matter term, i.e.  $T_{ab}^{DM}\equiv T^{int}_{ab}$. We
get
\begin{eqnarray}
    T_{ab}^{DM}=u_a u_b \sqrt{2 \alpha \rho_{M,0} \rho_{DE,0}}\,\,a^{-3}\equiv \rho_{DM} u_a u_b ,
\end{eqnarray}
According to the Raychaudhuri equation, for the change of the horizon area we have
\begin{equation}
\delta S = \int_H  \theta d\xi dS =-\int_H  \xi R_{ab}k^a\,k^b d\xi dS,
\end{equation}
in which $\theta$ is known as the expansion/contraction scalar. Finally, for the first law of thermodynamics we obtain
\begin{eqnarray}
    \delta (Q^M+Q^{DE}+Q^{int}) = T d\mathbb{S},
\end{eqnarray}
where $\mathbb{S}$ is the entropy and we note that $T$ is the
temperature and it reads $T=\kappa/2\pi$. Using the
Bekenstein-Hawking relation between the entropy and the horizon
area $\mathbb{S}=S/4$, and with the help of the above relations we
get
\begin{equation}
\int_H  \xi \left( T^M_{ab} +T^{DE}_{ab}+T^{int}_{ab}\right)k^ak^b d\xi dS= \frac{1}{8\pi G} \int_H  \xi R_{ab}k^a k^b d\xi dS.
\end{equation}
From the last equation it is evident that one has
\begin{equation}
\frac{1}{8 \pi G} \left(R_{ab}\,k^a\,k^b\right)=\left(T^M_{ab}
+T^{DE}_{ab}+T^{int}_{ab}\right)k^ak^b,
\end{equation}
which is valid for all null vector $k^a$. The last equation can be further rewritten as
\begin{eqnarray}
   8 \pi G T^{tot}_{ab}=R_{ab}+\zeta g_{ab},
\end{eqnarray}
with $\zeta$ being some function.  In addition we have defined the total energy-momentum tensor to be
\begin{eqnarray}
    T_{ab}^{tot}=T^M_{ab} +T^{DE}_{ab}+ T_{ab}^{DM}.
\end{eqnarray}
If we apply the covariant derivative to and we further utilize the contracted Bianchi identity, we obtain:
\begin{eqnarray}
\nabla_{b}(R_{ab}+\zeta g_{ab})=0\,\, \Rightarrow     -\nabla_{b}\left(\frac{R}{2}\right)=\partial_{b}\zeta.
\end{eqnarray}
We have also assumed $\nabla_b T^{tot}_{ab}=0$. From the last equation it is easy to see that we get for the function $\xi$
\begin{eqnarray}
    \xi=-\left( \frac{R}{2} \right)+C.
\end{eqnarray}
In our case we can fix the constant $C$ to zero. But we further
define
\begin{eqnarray}
    T^{DE}_{ab}=-\frac{\Lambda}{8 \pi G} g_{ab}.
\end{eqnarray}
In this way we obtain the Einstein field equations as a
thermodynamics equation of state with dark matter
\begin{eqnarray}
 R_{ab}-\frac{1}{2}R g_{ab}+\Lambda g_{ab}=8 \pi G \left( T^M_{ab} +T_{ab}^{DM}\right).
\end{eqnarray}
One should keep in mind that the dark matter is only an apparent
effect and obtained via the interaction of dark energy and matter.
With this equation in hand, we can apply to find the cosmological
model of the universe. Let us now use the Einstein equation with
dark matter and extend our discussion of the cosmological setup.
Assuming the background spacetime to be spatially homogeneous and
isotropic, which is given by the Friedmann-Robertson-Walker (FRW)
metric
\begin{equation}
ds^2=-dt^2+a^2\left[\frac{dr^2}{1-kr^2}+r^2(d\theta^2+\sin^2\theta
d\phi^2)\right],
\end{equation}
where we can further use $R=a(t)r$, $x^0=t, x^1=r$, the two
dimensional metric $ h_{\mu \nu}$. Here $k$ denotes the curvature
of space with $k = 0, 1, -1$ corresponding to flat, closed, and
open universes, respectively. The dynamical apparent horizon, a
marginally trapped surface with vanishing expansion, is determined
by the relation $h^{\mu
\nu}(\partial_{\mu}R)\,(\partial_{\nu}R)=0$. A simple calculation
gives the apparent horizon radius for the FRW universe
\begin{equation}
\label{radius}
 R=ar= {1}/{\sqrt{H^2+ {k}/{a^2}}}.
\end{equation}
For the matter source in the FRW universe, we shall assume a
perfect fluid described by the stress-energy tensor
\begin{equation}\label{T}
T_{\mu\nu}=(\rho_i+p_i)u_{\mu}u_{\nu}+p_ig_{\mu\nu}.
\end{equation}
this leads to the continuity equation have assumed several matter
fluids with a constant equation of state parameters $\omega_i$ and
continuity equations we have the expression for densities $\rho_i=\rho_{i 0}
a^{-3 (1+\omega_i)}$ (this includes baryonic matter $\omega_i=0$,
radiation $\omega_i=1/3$, dark matter $\omega_i=0$ and dark energy
$\omega_i=-1$). with $H=\dot{a}/a$ being the Hubble parameter. If
we rewrite the Einstein equations as
\begin{eqnarray}
 R_{ab}=8 \pi G \left( T^{tot}_{ab}-\frac{1}{2}g_{ab} T^{tot}\right).
\end{eqnarray}
From the Einstein field equations, we can obtain the Friedmann's
equation, namely
\begin{equation}
\frac{\ddot{a}}{a}=- \left(\frac{4 \pi G }{3}\right)\sum_i
\left(\rho_i+3p_i\right). \label{Aaddot}
\end{equation}
where we have the expression for densities $\rho_i=\rho_{i 0}
a^{-3 (1+\omega_i)}$. Eq. \eqref{Aaddot} becomes
\begin{align}
\frac{\ddot{a}}{a} =& - \left(\frac{4 \pi G }{3}\right)\sum_i
\left(1+3\omega_i\right) \rho_{i 0} a^{-3 (1+\omega_i)}.
\label{2Aaddot}
\end{align}
Next, by multiplying $2\dot{a}a$ on both sides of Eq.
\eqref{2Aaddot}, and by integrating we obtain
 \begin{align}
  \left(\frac{\dot{a}}{a}\right)^2+\frac{k}{a^2}=  \frac{8\pi G }{3} \sum_i \rho_{i0} a^{-3 (1+\omega_{i})}
 \end{align}
 or
\begin{align}
H^2+\frac{k}{a^2} =  \frac{8\pi G}{3}\sum_i \rho_{i0} a^{-3
(1+\omega_{i})}  \label{Fried01}
\end{align}
We further assume, the energy-momentum tensor of
each component is in the form of the perfect fluid,
\begin{eqnarray}
    T_{ab}^{i}=(\rho_{i}+p_{i})u_{a}u_{b}+p_{i}g_{ab}
\end{eqnarray}
where  $i=DM, DE, M$ and $u^{a}$ is the 4-velocity
of the prefect fluid with $g^{ab} u_{a} u_{b}=-1$. For the pressureless dark matter with $p_{DM}=0$, we have
\begin{eqnarray}\label{TDM1}
   {{T^a}_{b}}^{DM}=\rho_{DM}u^{a}u_{b}.
\end{eqnarray}
Comparing with the equation
\begin{eqnarray} \label{TDM2}
    {{T^{a}}_b}^{DM}= u^{a} u_{b}\sqrt{2 \alpha \rho_{M,0} \rho_{DE,0} }\,\,a^{-3}.
\end{eqnarray}
Equating Eqs.  (\ref{TDM1}) and  (\ref{TDM2}) we
finally get
\begin{eqnarray}
    \rho_{DM}=\sqrt{2 \alpha \rho_{M,0} \rho_{DE,0} } (1+z)^{3}.
\end{eqnarray}
Using the continuity equation for the matter, we
have $\rho_{M}= \rho_{M,0} (1+z)^3$, while for the cosmological
constant (DE) the energy density is constant, namely
$\rho_{DE}=\rho^{0}_{DE}$, in the general case we define
\begin{equation}
    \rho_{DM}=\sqrt{2\alpha \rho^0_{M} \rho^0_{DE}} (1+z)^3,
\end{equation}
or
\begin{equation}
 \rho_{DM}=\rho^0_{DM} \,(1+z)^3,
\end{equation}
that it we obtain the standard Friedmann equation
\begin{eqnarray}\notag
    \left(\frac{\dot{a}}{a}\right)^2&=&\frac{8\pi G}{3}\Big(\rho_{R,0}\, a^{-4}+\rho_{M,0}a^{-3}+\rho_{DM,0}a^{-3}\\
    &+&\rho_{\rm curv} a^2+\rho_{DE,0}\,a^0\Big),
    \end{eqnarray}
In addition, we have the spatial curvature density
\begin{eqnarray}
    \rho_{\rm cur}=\frac{3 k}{8 \pi G a^2}.
\end{eqnarray}
If we use the critical density
\begin{eqnarray}
    \rho_{\rm crit}=\frac{3 H_0^2}{8 \pi G },
\end{eqnarray}
we get the equation in terms of the density parameters or in terms
of redshift $z$ as follows
\begin{equation}\label{EofZ}
  E^2(z)=\Omega_{R,0}(1+z)^4+ \Omega_{M,0}^{\rm tot}(1+z)^{3}
  +\Omega_{\rm curv} (1+z)^2+\Omega_{DE,0}.
  \end{equation}
where $ \Omega_{M,0}^{\rm tot}=\Omega_{M,0}+\sqrt{2\,\alpha\,
\Omega_{M,0} \Omega_{DE,0} }$. This result coincides with Eq. (21)
in the last section.
%%%%%%%%%%%%%%%%%%%%%%%%%%%%%%%%%%%%%%%%%%%%%%%%%%%%%%%%%%%%%%%%%%%%%%%%%%%%%%%%%%%%%%%%%%
\section{Correspondence with Yukawa cosmology and recovering MOND}\label{sectV}
\subsection{Correspondence with Yukawa cosmology}
In this section, we would like to elaborate more on the nature of
dark energy. Einstein famously introduced the cosmological
constant into the field equation $G_{\mu \nu}+\Lambda g_{\mu
\nu}=8\pi G T_{\mu \nu}/c^4$ to explain the expansion of the
Universe. Dark energy is commonly linked to represent the energy
density of the vacuum \cite{Carroll:2000fy}; hence, it is just a
constant. Specifically, the energy density can be calculated by
integrating the vacuum fluctuation energies using $
    \rho^{\rm vac}_{DE} \simeq  \Lambda_{\rm cut}^4 $
where we integrate up to a certain ultraviolet momentum cutoff
$k_{\rm max}$, we get $\rho^{\rm vac}_{DE}\sim \hbar k^4_{\rm
max}$ \cite{Carroll:2000fy}. That is the origin of the famous
discrepancy, namely, when one compares the theoretical value with
the observed value, that gives a huge discrepancy of the order
$\sim 10^{120}$. Speculation regarding a potential symmetry leads
to a vanishing vacuum energy. The problem surrounding the
cosmological constant remains unsolved and continues to challenge
researchers. Certainly, alternative concepts have been proposed to
elucidate the cosmological constant. In this context, we will
present an argument for the involvement of massive gravitons in
clarifying the nature of the cosmological constant. Moreover, we
will draw attention to a correlation with Yukawa cosmology, the
subject of recent investigation in \cite{Jusufi:2023xoa,
Gonzalez:2023rsd}. At large distances, one can obtain a
Yukawa-like potential
\begin{eqnarray}
    \Phi = -\frac{G M m}{r}\left(1+\alpha
    \exp{(-r/\lambda)}\right),
\end{eqnarray}
with the wavelength of massive graviton reads \cite{Visser:1997hd}
\begin{equation}
\lambda=\frac{\hbar} {m_g c}.
\end{equation}
It is interesting to note that the Yukawa potential has been
obtained in modified theories of gravity such as the $f(R)$
gravity \cite{Capozziello:2009vr,Benisty:2023qcv}). It was shown
that the dark matter can be viewed as an apparent effect, namely,
dark matter is obtained via the long-range force modification in
terms of the following equation
\cite{Gonzalez:2023rsd,Jusufi:2023xoa}. In particular, one can
obtain effectively the $\Lambda$CDM using
\cite{Gonzalez:2023rsd,Jusufi:2023xoa}
\begin{eqnarray}
\Omega_{DM,0}^{\Lambda\text{CDM}}=\sqrt{2\Omega^{\Lambda\text{CDM}}_{M,0}
\Omega^{\Lambda\text{CDM}}_{DE,0}}, \label{OB0def}
\end{eqnarray}
where
\begin{equation}
\Omega^{\Lambda\text{CDM}}_{DE,0} \sim \frac{c^2 \alpha}{\lambda^2
H_0^2}.
\end{equation}
We can therefore make the correspondence if we define the dark
energy density in the Padmanbhan universe as
\begin{equation}
  \Omega^{\rm Padmanbhan}_{DE,0} \sim \frac{c^2}{\lambda^2 H_0^2}.
\end{equation}
In other words, there is an equivalence with the Yukawa cosmology
\cite{Jusufi:2023xoa} with two parameters: $\lambda$ that
specifies the dark energy density and it is related to the
gravitons mass and $\alpha$ that specifies the interaction between
matter and dark energy. We can further write the energy density of
the cosmological constant contribution in terms of the graviton
mass. Since we found a correspondence with Yukawa cosmology, one
can see that the parameter $\alpha$ appears in the modified
Newton's potential and, therefore, Newton's law in large
distances. The quest for a viable theory grounded in a healthy and
well-defined action that can integrate the concept of massive
gravitons remains an ongoing endeavour that has yet to reach a
conclusive resolution.  the parameter $\alpha$ plays a crucial
role in altering Newton's law. Yet, an intriguing question emerges
regarding the fundamental origin of this parameter. We shall
present compelling arguments linking the modified entropy to
adjusted gravity, with $\alpha$ emerging as a consequence of
entropy modification. In some deep sense, $\alpha$ could
potentially be influenced by the entanglement entropy arising from
the intricate interplay between baryonic matter and the
fluctuations in the gravitational field (gravitons). One can
already see a hint for this if we  use
$\Omega_{\Lambda,0}^{\Lambda CDM}=\rho_{DE,0}/\rho_{\rm
crit}$, where $\rho_{\rm crit}=3H_0^2/(8\pi G)$ and
$\rho_{DE,0}=\Lambda c^2/(8 \pi G)$ one can obtain the
cosmological constant in terms of $\alpha$, as follows
\begin{eqnarray}
    \Lambda \sim \frac{3 \alpha }{\lambda^2}.
\end{eqnarray}
In this way, we get for the energy density of dark energy
\begin{eqnarray}
    \rho_{DE} \sim \frac{3\, \alpha \,c^2}{8 \pi G
    \lambda^2},
\end{eqnarray}
which has the form of a holographic dark energy expression for the
cosmological constant. This signals that $\alpha$ fundamentally
measures the entanglement between the gravitons and matter fields.

\subsection{Recovering MOND}
In this section, we would like to point out that the parameter
$\alpha$ that modifies the Newton's law in the large distances,
can explain theoretical origin of the MOND theory. To see this,
let us rewrite Eq. \eqref{frw} as follows
\begin{equation}
 \frac{\ddot R}{R}=-\frac{4\pi G}{3}  \left[\sum_i  \rho_{i}(1 + 3 \omega_i )+\sqrt{2\alpha \rho_M \rho_{DE}}\right],
\end{equation}
where we have restored the Newton's constant $G$. This equation can also be written as
\begin{equation}
 m \ddot R=-\frac{G\,m}{R^2} \left[\frac{4\pi  R^3}{3}\sum_i  \rho_{i}(1 + 3 \omega_i )+\frac{4\pi R^3}{3} \sqrt{2\alpha \rho_M \rho_{DE}}\right].
 \label{(29)}
\end{equation}
Using the fact that the Komar mass [similar to Komar energy with $p_i=\omega_i \rho_i$] reads
\begin{eqnarray}
    M=\frac{4\pi  R^3}{3}\sum_i  \rho_{i}(1 + 3 \omega_i ),
\end{eqnarray}
and if we consider only matter $[\omega_i=0$] and dark energy $\omega_i=-1$] (we shall neglect radiation contribution), we get
\begin{equation}
 m \ddot R=-\frac{G\,m}{R^2} \left[\frac{4\pi  R^3}{3}(\rho_M-2\rho_{DE})+\frac{4\pi R^3}{3} \sqrt{2\alpha \rho_M \rho_{DE}}\right].
\end{equation}
For Newton's law in cosmological scales, we therefore get
\begin{equation}
    \vec{F}=m\, \vec{a}=-\left[\frac{G\,\,m\,(M-M_{DE})}{R^2}+\frac{G\,\,m}{R^2} V\,\sqrt{2\alpha \rho_M \rho_{DE}}\right]\vec{r_0},
\end{equation}
where $\vec{r_0}$ is some unit vector, and $M=4\pi R^3 \rho_M/3$
(although in general $M$ depends on $R$) along with $M_{DE}=4 \pi
R^3 (2\rho_{DE})/3$. On cosmological scales, the dark energy mass
$M_{DE}$ dominates over baryonic mass $M$, implying a negative
sign which means an expanding universe. Conversely, on galactic
scales, the baryonic mass dominates and the total force is toward
the galactic center. Nevertheless, the presence of the second term
which depends on $\alpha$, yields an additional contribution from
the long-range gravity modification; such a term can be attributed
to the dark matter having a mass
\begin{eqnarray}
    M_{DM}=\frac{4\pi  R^3}{3}\sqrt{2\alpha \rho_M \rho_{DE}},  \label{(33)}
\end{eqnarray}
we can restore Newton's law of gravity
\begin{eqnarray}
    \vec{F}=-\frac{G\,m\,(M-M_{DE}+M_{DM})}{R^2}\vec{r_0}.\label{(34)}
\end{eqnarray}
provided there is an extra mass term due to the dark matter. However, there is another way of  rewriting Eq. \eqref{(29)} as follows
\begin{eqnarray}
    a=\frac{G\,\,(M-M_{DE})}{R^2}+\sqrt{\alpha \left(\frac{G\,\,M}{R^2}\right)  \left(\frac{G\,\,M_{DE}}{R^2}\right)}
\end{eqnarray}
Let us define the acceleration due to the matter and the dark energy enclosed in that region of volume $V$ as
\begin{eqnarray}
    a_M=\frac{G\,\,M}{R^2}\,\,\,\,\,\text{and}\,\,\,\,\,  a_{DE}=\frac{G\,\,M_{DE}}{R^2},
\end{eqnarray}
we obtain the total acceleration
\begin{eqnarray}
    a=a_M-a_{DE}+\sqrt{\alpha a_M a_{DE}}.
\end{eqnarray}
The last equation is the acceleration in MOND theory.

\begin{itemize}
    \item Cosmological scales
\end{itemize}
In cosmological scales, as we pointed out the dark energy
dominates over baryonic matter [in general it also should depend
in the scale distance $M=M(R)$], and for cosmological scales  $R
\sim 10^{26} \rm {m}$, $a_M \to 0$, or $a_M<<a_{DE}$ and, from the
lat equation we get $a=-a_{DE}$.  It is clear that the negative
sign explains the expanding universe due to dark energy.  One can
further check that
\begin{eqnarray}
    a_{DE}=\frac{G\,\,M_{DE}}{R^2} \sim \frac{4 \pi G \rho_{DE} R}{3},
\end{eqnarray}
where for the dark energy density we have
\begin{eqnarray}
    \rho^{\rm cosmology}_{DE} \sim \frac{3\, \alpha \,c^2}{8 \pi G
    \lambda^2_{\rm cosmology}},
\end{eqnarray}
hence we get
\begin{eqnarray}
    a_{DE} \sim \frac{\alpha c^2 R}{2 \lambda^2_{\rm cosmology}} \sim 10^{-10} \rm{m/s^2}.
\end{eqnarray}
The last result gives the acceleration of the universe. Here we
note that the value of $\lambda_{\rm cosmology}$ is of the order
of the observable radius of the universe, i.e., $\lambda_{\rm
cosmology}\sim 10^{26}$ m (see from observation constrains
reported in \cite{Jusufi:2023xoa,Gonzalez:2023rsd}). In addition,
we have used $\alpha \sim 0.41$ along with $R \sim 10^{26} \rm
{m}$, and consequently the energy density of dark energy can be
approximated to $\rho_{DE}\sim 7 \times 10^{-27}$ kg/m$^3$.

\begin{itemize}
\item Galactic scales
\end{itemize}
Let us now elaborate the case of galaxies where $R$ is of pc or
kpc order. However, near the galactic center in general the mass
term is a function of distance, i.e., $M(R)$. In other words, in
the galactic center, we know that baryoinic matter dominates
compared to the dark energy ($a_M>>a_{DE})$, hence we get
\begin{eqnarray}
  a=a_M+\sqrt{\alpha\, a_M\,a_{DE}}.
\end{eqnarray}
Further let us define $a_0$, by writing
\begin{equation}
    a_0 = \alpha\, a_{DE},
\end{equation}
to get a MOND-like form for the acceleration
\begin{eqnarray}
    a=a_M+\sqrt{ a_M a_0}.
\end{eqnarray}
The interesting part here is when one studies the large distances,
say when $R$ is of pc or kpc order. Hence let us take for the
outer part of the galaxy $R \sim 10^{17}$ m, we can try to compute
$a_0$ which gives
\begin{equation}
    a_0 = \alpha\, a_{DE} \sim \frac{4 \pi \alpha^2 G \rho_{DE} R}{3} \sim 10^{-20} \rm{m/s^2}!
\end{equation}
We have expected for $a_0$, as MOND predicts, $a_0\sim 10^{-10}
\rm{m/s^2}$. To address this inconsistency, let's initially
highlight that the energy density of dark energy on galactic
scales is given by
\begin{eqnarray}
    \rho^{\rm galaxy}_{DE} \sim \frac{ 3 \alpha \,c^2}{8 \pi G
    \lambda^2_{\rm galaxy}},
\end{eqnarray}
and rewrite $a_0$ in terms of the energy density as follows
\begin{eqnarray}
   a_0 \sim \frac{\alpha^2 \,c^2}{2
    \lambda^2_{\rm galaxy}}.
\end{eqnarray}
As we shall elaborate the energy density of dark energy depends on
the scale of measurements (observations) due its dependence on
$\lambda$. In some sense there exists a fundamental limitations on
measuring the dark energy density due to the uncertainty principle
when we measure $\lambda$.  To see this more clearly, let us go
back in Eq. (59) and after we set $\lambda = \Delta x$, and
$\Delta p= m_g c$, for graviton we get the momentum-position
uncertainty relation
\begin{eqnarray}
    \Delta x  \Delta p \sim \hbar.
\end{eqnarray}

Now lets see the implications of this relation. When $\lambda^{\rm
cosmology} \sim 10^{26}\, \rm{m}\,.$, i.e. in cosmological scales,
we have more uncertainty in position $\lambda^{\rm cosmology}
=\Delta x^{\rm cosmology} \sim 10^{26}$ m,  but more precision on
the momentum $\Delta p^{\rm cosmology}$, namely
\begin{eqnarray}
     \Delta p^{\rm cosmology} \sim \frac{\hbar}{\Delta x^{\rm cosmology}} \sim 10^{-60}\,\rm{N}\cdot\rm{s}\,.
\end{eqnarray}
This simply means that when we apply the uncertainty relation for the graviton entirety of the universe, our knowledge of position becomes more uncertain, but we gain greater precision in measuring momentum. Conversely, when focusing on galactic scales, the uncertainty in position decreases, approximately $\lambda^{\rm galaxy} =\Delta x^{\rm galaxy} \sim 10^{19}$ m. However, this reduction in position uncertainty is accompanied by an increase in momentum uncertainty, denoted as $\Delta p^{\rm galaxy}$, specifically we can see this from
\begin{eqnarray}
     \Delta p^{\rm galaxy} \sim \frac{\hbar}{\Delta x^{\rm galaxy}} \sim 10^{-53}\,\rm{N}\cdot\rm{s}\,.
\end{eqnarray}

In order to obtain a consistent result, we need $R \gtrsim  10^{15}$ m, namely
\begin{equation}
    a_0 \sim  \frac{ \alpha^2 \,c^2\, R}{2
    \lambda^2_{\rm galaxy}} \sim 10^{-10} \rm{m/s^2}.
\end{equation}
As we increase the scale of observations $R$, $\lambda$ will
increase and there is no sense to speak about $R>\lambda$. Our
findings align with recent observations related to the
applicability of Newton's law over substantial distances,
specifically employing wide binary systems as exemplified by
Hernandez and Chae \cite{Hernandez,Chae:2023prf}. Consequently, it
is evident that there are inherent constraints in accurately
measuring the graviton wavelength within the realm of cosmology.
As we elaborated, the inconsistency that emerges when comparing
analyses conducted on galactic and cosmological scales, can be
explained through the prism of uncertainty relations.
%%%%%%%%%%%%%%%%%%%%%%%%%%%%%%%%%%%%%%%%%%%%%%%%%%%%%%%%%%%%%%%%%%%%%%%%%%%%%%%%%%%%%%%%%
\section{Final remarks and conclusions}
The central and novel result we found in this paper is that by
adopting the emergence perspective of gravity, one can interpret
the dark matter as an effective/apparent phenomenon. In this
viewpoint the dark matter appears as a result of interaction
between dark energy and baryonic matter which modifies the numbers
of degrees of freedom in bulk. This approach leads naturally to
derive the $\Lambda$CDM model of cosmology. The second important
result is that, we applied the Jacobson's approach by taking into
account an interaction term between baryonic matter and dark
energy which plays the role of dark matter. We thus constructed
the modified Einstein field equations as an equation of state. The
Einstein field equations obtained in this method, contains energy
momentum tensor of apparent dark matter as well as dark energy.
Both methods are consistent and we confirm how dark matter
naturally emerges by including an interaction term between dark
energy and baryonic matter. Finally, we obtained the $\Lambda$CDM
model in cosmology and established a correspondence with Yukawa
cosmology. We then explored the relation and the role of massive
graviton in the dark sector. For the energy density of dark energy
we obtained a holographic dark energy expression which implies
that the interaction term $\alpha$ fundamentally measures the
entanglement between the gravitons and matter fields. Finally, a MOND-like expression was obtained and the role of uncertainty relation for the gravitons' wavelength was discussed in cosmological/galactic observations.

%%%%%%%%%%%%%%%%%%%%%%%%%%%%%%%%%%%%%%%%%%%%%%%%%%%%%%%%%%%%%%%%%%%%%%
\acknowledgments{The work of A.Sheykhi is supported by Iran
National Science Foundation (INSF) under grant No. 4022705.}

%%%%%%%%%%%%%%%%%%%%%%%%%%%%%%%%%%%%%%%%%%%%%%%%%%%%%%%%%%%%%%%%%%%%%%

\end{document}